\documentclass[prb,aps,twocolumn,showpacs,floatfix,preprintnumbers,amsmath,amssymb,superscriptaddress]{revtex4}

\usepackage{graphicx}
\usepackage{bm}
\usepackage{latexsym,color,natbib}

\def\tn {$T_{\rm N}$}
\def\ts {$T_{\rm S}$}

\def\mb {$\mu_{\rm B}$}

\newcommand{\tb}{TbFe$_{3}$(BO$_{3}$)$_{4}$}
\newcommand{\gd}{GdFe$_{3}$(BO$_{3}$)$_{4}$}

\newcommand{\figref}[1]{Fig.\,\protect\ref{#1}}

\begin{document}

\title{Magnetodielectric and magnetoelastic coupling in TbFe$_{3}$(BO$_{3}$)$_{4}$}
\author{U. Adem}
\affiliation{Leibniz Institute for Solid State and Materials Research (IFW) Dresden, Germany}
\author{L. Wang}
\affiliation{Leibniz Institute for Solid State and Materials Research (IFW) Dresden, Germany}
\author{D. Fausti}
\affiliation{Max Planck Group for Structural Dynamics-University of Hamburg-CFEL, Hamburg, Germany}
\author{W. Schottenhamel}
\affiliation{Leibniz Institute for Solid State and Materials Research (IFW) Dresden, Germany}
\author{P.H.M. van Loosdrecht}
\affiliation{Zernike Institute for Advanced Materials, University of Groningen,
Nijenborgh 4, 9747 AG Groningen, The Netherlands}
\author{A. Vasiliev}
\affiliation{Moscow State University, Moscow, 119992, Russia}
\author{L. N. Bezmaternykh}
\affiliation{Kirensky Institute of Physics, Siberian Division, Russian Academy of Sciences, Akademgorodok, Krasnoyarsk, 660036, Russia}
\author{B. B{\"u}chner}
\affiliation{Leibniz Institute for Solid State and Materials Research (IFW) Dresden, Germany}
\author{C. Hess}
\affiliation{Leibniz Institute for Solid State and Materials Research (IFW) Dresden, Germany}
\author{R. Klingeler}
\affiliation{Leibniz Institute for Solid State and Materials Research (IFW) Dresden, Germany}
\affiliation{Kirchhoff Institute for Physics, University of Heidelberg, Germany}

\date{\today}

\begin{abstract}

We have studied the magnetodielectric and magnetoelastic coupling in TbFe$_{3}$(BO$_{3}$)$_{4}$
single crystals by means of capacitance, magnetostriction and Raman spectroscopy measurements. The
data reveal strong magnetic field effects on the dielectric constant and on the macroscopic sample
length which are associated to long range magnetic ordering and a field-driven metamagnetic
transition. We discuss the coupling of the dielectric, structural, and magnetic order parameters
and attribute the origin of the magnetodielectric coupling to phonon mode shifts according to the
Lyddane-Sachs-Teller (LST) relation.
\end{abstract}

\maketitle

\newpage

\section{Introduction}

Many novel complex systems are investigated in the research on magnetoelastic and
magnetodielectric coupling \cite{Eerenstein,Cheongreview,Fiebigreview}. Among them,
rare-earth iron borates RFe$_{3}$(BO$_{3}$)$_{4}$ constitute a family of non-centrosymmetrric
oxides in which 4f-moments are embedded in a network of $3d$ Fe spins. These compounds crystallize
in trigonal R32 space group at high temperatures \cite{Joubert, Campa}. The members of the family
having rare-earth elements with ionic radii smaller than Sm undergo a structural phase transition
lowering the crystal symmetry from R32 to P3$_{1}$21.~\cite{Klimin, Fausti-PRB}. Based on single crystal
diffraction studies, it has been proposed that the structural transition results in an
antiferroelectric state \cite{Klimin}. Magnetic exchange interactions for R = (magnetic) Rare
Earths are complicated due to the 4f-3d interactions and the anisotropy of the 4f-moments which,
e.g., affects the direction of Fe-spin in the ordered phase.

Structurally, TbFe$_{3}$(BO$_{3}$)$_{4}$ exhibits edge sharing FeO$_{6}$ polyhedra which are
forming helicodial chains along the $c$ axis \cite{Campa}. Fe spins order
antiferromagnetically at \tn\ $\approx 39$ K along the helicoidal chains of Fe atoms with the
spins parallel to the hexagonal $c$ axis while Tb moments are polarized from Fe ordering
\cite{Popova,Volkov, Campa, Ritter}. When a magnetic field is applied along the easy axis $c$, a
field induced magnetic transition is observed at temperatures below $T_{\rm N}$, i.e. there is a
spin-flop of the Fe spins while the Tb moments become fully aligned.

So far R = Gd, Pr, Nd and Tb compounds have been studied in the context of multiferroics and
magnetoelectrics. In fact, the space groups of both high and low temperature phases are not polar
despite being non-centrosymmetric, therefore the family cannot be considered as multiferroic. On
the other hand, it has been shown for R = Gd, Pr, Nd and Tb that magnetic field induces electrical
polarization, marking the materials as magnetoelectric
\cite{ZvezdinGd,KadomtsevaPr,ZvezdinNd,ZvezdinTb}. Also, sizeable spin-lattice coupling was
demonstrated for some of these compounds using magnetostriction measurements
\cite{ZvezdinTb,Demidov}. Within the family, the dielectric properties were reported only for R =
Gd where the dielectric constant was found to decrease with decreasing temperature. The onset of
magnetic ordering causes an anisotropic anomaly in $\epsilon$. In the magnetically ordered phase
Gd 4f - Fe 3d magnetic interactions cause a spin reorientation transition which is coupled to the
dielectric constant via spin-lattice interaction.\cite{Fausti-PRB, Yen}. Application of magnetic
field strongly shifts the spin reorientation transition, giving rise to a 1$\%$ magnetodielectric
effect \cite{Yen}. Here, we present a detailed experimental study on the magnetodielectric and
magnetoelastic coupling in \tb . We have measured the dielectric constant and the magnetostriction
and we show by analysis of Raman spectroscopy data that the magnetodielectric coupling is mediated
by strain and caused by shifts in the corresponding transverse optical (TO) phonon modes via
Lyddane-Sachs-Teller (LST) relationship.

\section{Experimental}

The single crystals we used are from the same batch as in Ref.~\onlinecite{Popova}. The
capacitance of the sample was measured using a home-made insert and Andeen-Hagerling AH2500A
bridge in a 18 Tesla Oxford cryostat between 5 and 300 K. The measurement frequency was 1 kHz.
Magnetostriction measurements were carried out with a capacitance dilatometer\cite{Wang09} in
the same cryostat and with the same capacitance bridge. Our dilatometer applies the tilted plate
principle with the sample placed in the open center of a ring-like capacitor made from silver. It
enables to follow the length changes of the sample at constant temperature upon changing the
external magnetic field. The sample length was 1.23 mm. The field was changed quasi-statically from 0 up to 14 T and back with a rate of 0.01 T/s. The length changes of the
sample are calculated from the capacitance changes measured by a temperature stabilized
capacitance bridge Andeen-Hagerling with a resolution of 5*10$^{-7}$ pF. Thus, length changes of less
than 0.01 {\AA} can principally be resolved. Due to mechanical vibrations, etc., the resolution is
limited to 0.1-1 {\AA} in practice. The measurement frequency was 1 kHz. The field dependence of the
magnetization up to 15 T was studied in a home-built VSM magnetometer.~\cite{vsm}

The Raman measurements were performed in a backscattering configuration, using a three-grating
micro-Raman spectrometer (T64000 Jobin Yvon) equipped with a liquid nitrogen cooled charged
coupled device (CCD) detector. The frequency resolution was better than 2 cm$^{-1}$ for the
frequency region considered. The triple grating configuration allows the analysis of a broad
spectrum from 10000 cm$^{-1}$ down to 4 cm$^{-1}$. The sample was placed in an optical microscope
cryostat. The temperature was varied from 4 to 55 K, with a stability of 0.1 K. The scattering was
excited by the second harmonic light of a Nd:YVO4 laser (532 nm), focused to 50 $\mu$m$^{2}$  with
the power density on the sample kept below 0.1 mW/$\mu$m$^{2}$. The polarization was controlled
both on the incoming and outgoing beams giving access to all relevant polarization combinations.
The Raman measurements reported in the paper are performed with both the light's k-vector and the
polarization of the incoming and outgoing field in the BO$_{3}$ planes.

\section{Results and Discussion}

\begin{figure}[tbp]
\centering
\includegraphics[width=0.95\columnwidth]{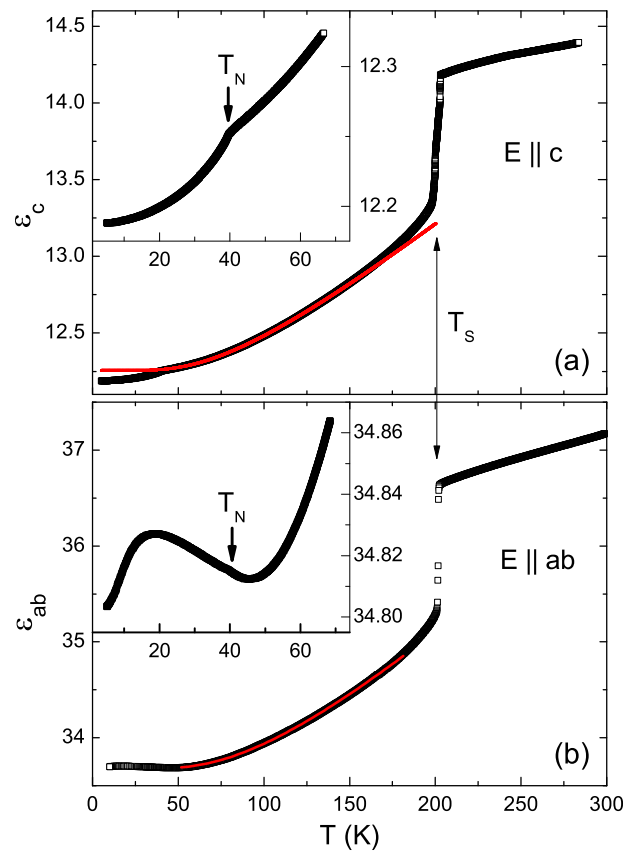}
\caption{Temperature dependence of the dielectric constant of TbFe$_{3}$(BO$_{3}$)$_{4}$, for
$E\parallel c$ (a) and $E \perp c$ (b). The insets highlight the data at low temperatures. The red
line show fits according to Barrett's function (see the text).} \label{fig1}
\end{figure}

Fig. 1(a) shows the temperature dependence of the dielectric constant measured along the $c$ axis.
In the whole temperature regime under study, $\epsilon_c$ decreases upon cooling. In addition, we
observe an abrupt decrease at the structural phase transition at \ts\ = 202 K. Also the magnetic
ordering of the Fe spins at \tn\ $\approx 39$ K causes an anomaly in the dielectric constant (see
the inset of Fig. 1(a)) which is however much smaller than that at \ts . The temperature
dependence of the dielectric constant measured perpendicular to the $c$ axis in general resembles
the behavior for $E\|c$ but there are subtle differences at low temperature (\figref{fig1}). We
again observe a discontinuity at \ts\ but unlike the case $E\parallel c$ there is only a very weak
anomaly at \tn\ (cf. inset of \figref{fig1}(b)). Instead, the data display a minimum above the
magnetic ordering temperature, i.e. around 50 K, below which the dielectric constant increases.
Upon further cooling, $\epsilon_{ab}$ decreases again below a broad maximum around 20 K.

\begin{figure}[tbp]
\centering
\includegraphics[width=0.95\columnwidth]{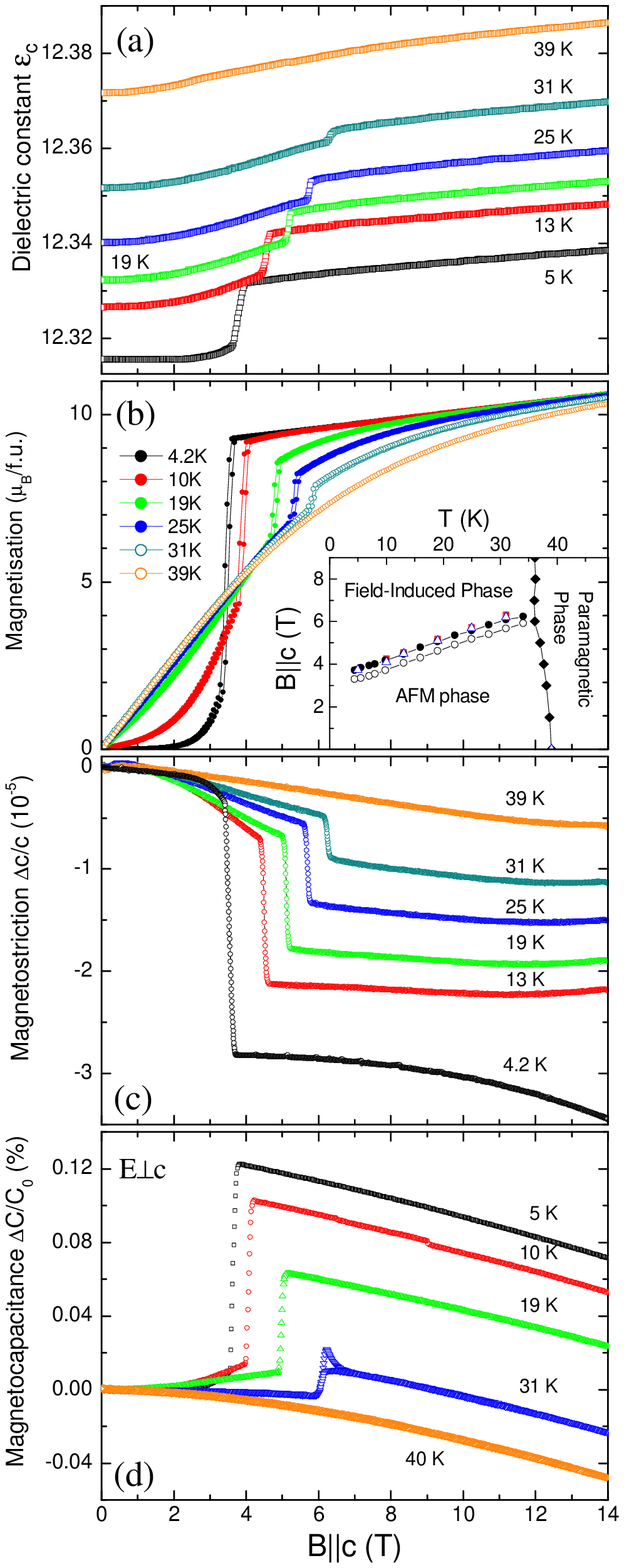}
\caption{ (a) The magnetic field dependence of the dielectric constant for E $\parallel$ $c$, B $\parallel$
$c$. (b) Magnetization (from Ref.~\onlinecite{Popova}), (c) Magnetostriction of the $c$ axis, and
(d) capacitance for E $\perp$ $c$, B $\parallel$ $c$. The inset displays the magnetic phase
diagram derived from dielectric constant, magnetization, magnetostriction and thermal expansion
studies.} \label{fig2}
\end{figure}

The effect of an external magnetic field on $\epsilon_{c}$  applied parallel to the $c$-axis is
shown in \figref{fig2}(a). At $T=5$ K, the capacitance is rather field independent for small
$B\|c$ but a sharp increase appears around 4 T. At higher fields, the capacitance increases nearly
linearly with a rather constant slope. The critical field of the jump in $\epsilon_c$ increases
upon heating and the size of the jump becomes smaller. Above \tn , i.e. in absence of long range
magnetic order, the jump $\Delta\epsilon$ vanishes. Note, that for all temperatures studied in the
magnetically ordered phase the behavior in the high-field phase is nearly independent of
temperature and the slope at high fields is very similar above \tn , too. The sharp anomaly in the
dielectric constant is associated to a previously found field induced magnetic transition (FIMT)
which is clearly visible in magnetization measurements (see. \figref{fig2}(b)).~\cite{Popova} The
magnetization data show a huge jump of roughly 9 \mb /f.u. at the FIMT, and  similar to the jump
in $\epsilon$ the anomaly $\Delta M$ decreases and shifts to higher fields upon heating. The field
induced magnetic transition is not only associated to jumps in $\epsilon_c$ and in the
magnetization but it is also accompanied by strong magnetostrictive effects. As seen in in Fig.
2(c), the $c$-axis exhibits a jump-like decrease at the critical field. Again, the jump reduces
upon heating and it is restricted to the magnetically ordered phase. In general, the $c$ axis
contracts in the whole field range studied. In Fig. 2(d), the magnetic field dependence of
$\epsilon_{ab}$  is shown, with the magnetic field again applied along the $c$ axis. We again
observe that the magnetic field dependence is small in both the low-field and the high-field
antiferromagnetic phase while the largest changes occur at the FIMT. However, unlike the
measurement for $\epsilon_{c}$, the dielectric constant decreases with magnetic field in the field
induced phase. The phase diagram shown in the inset summarizes the transition fields of the FIMT
vs. temperature as derived from dielectric constant, magnetization, magnetostriction and specific
heat.~\cite{Popova}. We note that sharp changes in magnetostriction and electrical
polarization at $T=4.2$\,K were reported recently in pulsed field studies, too.~\cite{ZvezdinTb,
Demidov} The reported anomalies in the magnetostriction are however about 2 times smaller than
found with our quasi-static set-up which might be associated to the different experimental
approach, i.e. pulsed magnetic field and usage of a piezoelectric sensor glued to the sample. In
addition, there are also qualitative differences to our data, i.e. in $\Delta L_a/L_a$ at $B||c >
B_c$ and pronounced hysteresis effects in the magnetostriction and the electrical polarization.

To summarize the experimental data in \figref{fig2}, at the FIMT \tb\ displays a magnetically
driven phase transition which is associated to sharp jumps in (1) the dielectric constant, (2) the
magnetization, and (3) in the length of the $c$-axis. It has been shown earlier for the magnetic
degrees of freedom that, at the FIMT, there is a spin-flop of the Fe spins accompanied by a
magnetic flip of the Tb moments.~\cite{Popova}

\begin{figure}[tbp]
\centering
\includegraphics[width=0.7\columnwidth]{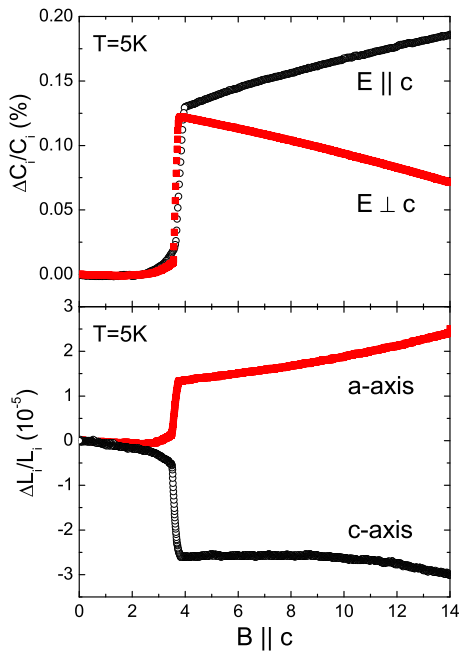}
\caption{(a) Magnetocapacitance parallel and perpendicular to the $c$-axis, and (b)
magnetostriction along the $a$- and $c$-axis, respectively, of \tb , at $T=5$ K, for magnetic
fields applied along the $c$-axis.} \label{fig3}
\end{figure}

The pronounced magnetoelastic coupling and the resulting jump of the sample length at the FIMT
might suggest that the observed step in the dielectric constant is only associated to the
magnetostriction, i.e.
\begin{equation}\label{cap}
C = \epsilon_{0}\epsilon\frac{A}{d}
\end{equation}
where $\epsilon$$_{0}$ is the permittivity of the free space, $\epsilon$ is the dielectric
constant, $A$ is the area of the contacts, and $d$ is the thickness of the sample. In general,
a magnetic field induced jump in the sample length implies a capacitance anomaly which can not be
discussed in terms of the magnetodielectric effect. This scenario can however be clearly ruled out
if the anisotropy of the magnetostriction and magnetocapacitance is considered (cf.
\figref{fig3}). If a magnetic field is applied along the $c$-axis, the length of the $c$-axis and
the $a$-axis sharply shrinks and increases, respectively. Such an opposite sign of the length
jumps is expected due to at least partial volume conservation of the magnetoelastic distortion. In
contrast, both jumps in $\epsilon_c$ and $\epsilon_{ab}$ exhibit the same sign, i.e.
magnetostriction has only a minor effect on the observed field dependence of $\epsilon$ and cannot
explain the changes in $\epsilon$($B$). Therefore, the anomalies clearly point to a direct
magnetodielectric coupling.

As discussed above, the general behavior of the dielectric constant vs. temperature is rather
isotropic while there is a weak difference between $\epsilon$$_{c}$(T) and $\epsilon$$_{a}$(T) at
low temperatures. Interestingly, despite the fact that in \gd\ the magnetic structure in the
ordered phase is different compared to \tb , the anomaly at \tn\ as well as the low temperature
behavior $\epsilon_c$($T$) and $\epsilon_{ab}$($T$) is roughly similar in both
compounds.~\cite{Yen} In particular, both materials exhibit an increase of $\epsilon_{ab}$ below
\tn\ which in \gd\ is truncated by a spin-reorientation transition. For
GdFe$_{3}$(BO$_{3}$)$_{4}$, the upturn was suggested to originate from the polarization of Gd
spins and the vicinity of a spin-reorientation transition \cite{Yen}. In contrast, there is no
spin-reorientation in TbFe$_{3}$(BO$_{3}$)$_{4}$ but as well the specific heat as the
magnetization data exhibit a broad Schottky-like anomaly at the temperature where there is the
peak in $\epsilon_{ab}$. Recently, the anomalies of the specific heat and the magnetization were
explained in terms of temperature-driven population of the ground state of Tb ion split by the
internal field of Fe spins ~\cite{Popova} which suggests a similar, i.e. Schottky-like, scenario for the broad
peak in in $\epsilon_{ab}$.

In the following we will discuss the origin of the observed magnetodielectric coupling. In
general, the dielectric constant is related to optical phonon frequencies via the
Lyddane-Sachs-Teller relationship:
\begin{equation}\label{eq3}
\epsilon_{0}=\frac{\omega_{L}^{2}}{\omega_{T}^{2}} \epsilon_{\infty}
\end{equation}
In this equation $\omega_{L}$ and $\omega_{T}$ are the long wavelength longitudinal and transverse
optical-phonon mode frequencies, respectively. $\epsilon_{0}$ is the static dielectric constant,
i.e. the dielectric constant at zero frequency, and $\epsilon_{\infty}$ is the optical dielectric
constant. Following the early literature on BaMnF$_{4}$ \cite{Fox}, MnO \cite{SeehraMnO} and
MnF$_{2}$ \cite{SeehraMnF2}, we have tried to relate the dielectric constant we measured to the
relevant TO phonon modes. In order to do so, we have used a modified Barrett equation as described
in the aforementioned references as follows:
\begin{equation}
    \epsilon(T)=\epsilon(0)+A /[exp(\hbar \omega_{0}/k_{B}T)-1]
\end{equation}
In this equation, $A$ is a coupling constant and $\omega$$_{0}$ is the mean frequency of the final
states in the lowest lying optical phonon branch. By fitting our data as indicated by the red
lines in \figref{fig1} we obtained $A=33.6$, $\omega_{0}=206$ cm$^{-1}$ for $\epsilon$$_{ab}$ and
$A=12.2$, $\omega_{0}=177$ cm$^{-1}$ for $\epsilon$$_{c}$. We note the tranverse optical mode
that we obtained from the Barrett function fits corresponds to an average of all the contributing
transverse optical modes in the respective direction \cite{SeehraMnF2}.

\begin{figure}[tbp]
\centering
\includegraphics[width=1\columnwidth]{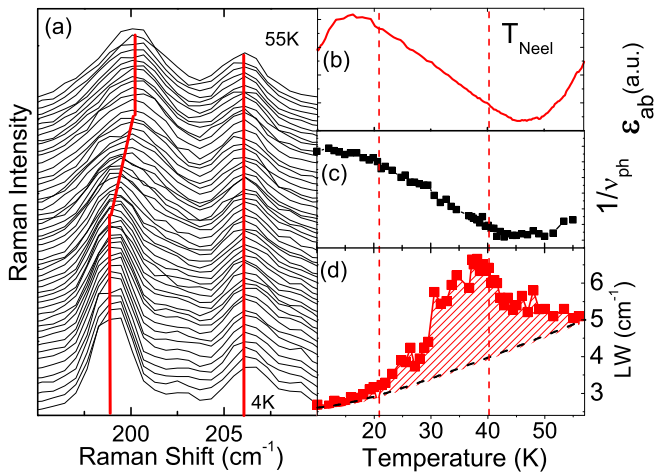}
\caption{Comparison between the behaviour of the phonon mode and the dielectric constant. (a)
Temperature dependence of the phonon frequency responsible of the anomaly measured in the
dielectric constant. (b) Dielectric constant measured perpendicular to the c axis. (c) Inverse of
the phonon frequency as a function of temperature. (d) Line-width of the phonon mode associated to
the dielectric constant anomalous behaviour.} \label{Raman}
\end{figure}

We have done Raman spectroscopy measurements to look for direct evidence of spin-lattice coupling.
According to the previous reports \cite{Fausti-PRB, deAndres}  there are 59 transverse optical modes that correspond to $\epsilon_{ab}$ i.e. phonon modes corresponding to a TO mode propagating perpendicular to the $c$ axis. We have checked the temperature dependence of these modes and observed that only two low lying ones shift their frequency at low temperatures: the first one at 200 cm$^{-1}$, the second one at 260 cm$^{-1}$. The modes at other frequencies do not significantly change their frequency. We show in \figref{Raman}(a) two representative low-lying phonon modes around 200 cm$^{-1}$. Upon cooling below the magnetic ordering temperature, one of the modes with the frequency 206 cm$^{-1}$ does not shift with changing temperature while the frequency of the lower lying transverse phonon mode at 200 cm$^{-1}$ is shifting remarkably at \tn. Its shift provides direct evidence for the
spin-lattice coupling since exchange striction stretches or elongates the bonds and hence shifts
the phonon frequencies when magnetic ordering takes place.

The anomalous softening of the phonon mode reported in \figref{Raman}(a) is a possible reason for
the behavior of the dielectric constant. Below the antiferromagnetic ordering temperature \tn ,
the behavior of the dielectric constant perpendicular to $c$ (as discussed in the previous
section) is rationalized in the light of the Raman response. As suggested by the
Lyddane-Sachs-Teller (LST) equation the dielectric constant is inversely proportional to the
frequency of the TO-phonon mode, associated to it. We note that usually LO-modes and
$\epsilon_{\infty}$ are assumed as temperature independent \cite {Samara, SeehraMnF2}.
\figref{Raman} summarizes the qualitative agreement between temperature dependence of the
dielectric constant (b) and the inverse of the phonon frequency (c). Remarkably, the measured
changes of the dielectric constant correspond very well to a decrease of the phonon frequency as
described by the LST equation. We note however that in order to have a quantitative
agreement, one needs to add the contributions from the all the relevant symmetry allowed TO
modes.

In addition to this, the data imply a large anomalous broadening of the phonon mode at the
magnetic transition itself. This is shown in \figref{Raman}(d) where the phonon line-width as a
function of temperature across the antiferromagnetic ordering temperature is presented. The black
dashed line describes the expected decrease of the phonon lifetime as a function of temperature as
if it was purely due to the usual phonon anharmonicity. 
The (red) dashed area in \figref{Raman}(d) reveals that in the temperature region close to the
magnetic transition another decaying channel is available for the transverse optical mode. In the
vicinity of the Neel temperature the phonon broadening deviates much stronger from the expected
behavior suggesting that a phonon-magnon coupling channel becomes active at the magnetic ordering
temperature. 
This strongly suggests that the additional decaying channel reducing the phonon lifetime is not
directly coupled to the magnetic order parameter but rather to its fluctuations.

In conclusion, we have demonstrated magnetodielectric coupling in TbFe$_{3}$(BO$_{3}$)$_{4}$ using
capacitance measurements and revealed the correlation between spin-phonon coupling and dielectric
constant. We have shown that magnetodielectric coupling cannot be explained by the
magnetostrictive effects but it occurs via shifting of the associated optical phonon modes.
The magnetic field induced phase transition the magnetic field induced phase transition is
associated to sharp changes of not only the magnetic but also dielectric and structural degrees of
freedom, i.e. the spin-flop of Fe spins is accompanied not only by full polarization of the Tb
moments but also to a significant distortion probably of the FeO$_6$ octaedra and a jump in the
dielectric constant.

\begin{acknowledgments}We thank K.~Leger and S. Ga{\ss} for technical support. Work was supported by the DFG through HE
3439/6 and 486 RUS 113/982/0-1.
\end{acknowledgments}

\end{document}